\documentclass[superscriptaddress,aps,twocolumn,pre]{revtex4}
\usepackage{graphicx}
\begin{document}
\title{{\bf Conditional Probability as a Measure of Volatility Clustering in Financial Time Series}}
\author{Kan Chen}
\affiliation{Department of Computational Science, Faculty of Science,
National University of Singapore, Singapore 117543}
\author{C. Jayaprakash}
\affiliation{Department of Physics, Ohio State University, 191 West Woodruff Avenue,
Columbus, Ohio 43210, USA}
\author{Baosheng Yuan}
\affiliation{Department of Computational Science,
Faculty of Science, National University of Singapore, Singapore 117543}

\date{\today }

\begin{abstract}

In the past few decades considerable effort has been expended in
characterizing and modeling financial time series. A number of
stylized facts have been identified, and volatility clustering or
the tendency toward persistence has emerged as the central
feature. In this paper we propose an appropriately defined
conditional probability as a new measure of volatility clustering.
We test this measure by applying it to different stock market
data, and we uncover a rich temporal structure in volatility
fluctuations described very well by a scaling relation. The scale
factor used in the scaling provides a direct measure of volatility
clustering; such a measure may be used for developing techniques
for option pricing, risk management, and economic forecasting. In
addition, we present a stochastic volatility model that can
display many of the salient features exhibited by volatilities of
empirical financial time series, including the behavior of
conditional probabilities that we have deduced.

\end{abstract}

\maketitle


\section{Introduction}

Forecasting from time series data necessarily involves an attempt
to understand uncertainty; volatility or the standard deviation is
a key measure of this uncertainty and is found to be time-varying
in most financial time series. The seminal work of
Engle \cite{Engle}, that first
treated volatility as a process rather than just a number to
estimate, led to tremendous efforts in devising dynamical
volatility models in the last two decades. These are of great
importance in a variety of financial transactions including option
pricing, portfolio and risk management. Excess volatility (well
beyond what can be described by a simple Gaussian process) and the
associated phenomenon of clustering \cite{benoit,Fama} are believed to be 
the key factors
underlying many empirical statistical properties of asset prices,
characterized by a few key ``stylized facts''
\cite{intro1,intro2,intro3,intro4} described later. A good measure
of volatility clustering (roughly speaking, large and small
changes in asset prices are often followed  by large and small
changes respectively) is thus important for understanding
financial time series and for constructing and validating a good
volatility model. The most popular characterization of volatility
clustering is the correlation function of the instantaneous
volatilities evaluated at two different times, which shows
persistence up to a time scale of more than a month. It has also
been established that there is link between asset price volatility
clustering and persistence in trading activity (for an extended
empirical study on this, see Ref.~\cite{Plerou}). However, the
underlying market mechanism for volatility clustering is not
clear. The aim of our paper is not to elucidate the mechanism for
volatility clustering, but to introduce a more direct measure of
it. Specifically, we propose that the {\em conditional}
probability distribution of asset returns $r$ over a period $T$
(given the return, $r_p$, in the previous time period )can be
fruitfully used to characterize clustering. This is a direct
measure based on return over a time lag instead of instantaneous
volatility and we believe is more relevant to volatility
forecasting. We analyze stock market data using this measure, and
we and have found that the conditional probability can be well
described by a scaling relation: $P(r|r_p) = \frac{1}{w(r_p)}
f(r/w(r_p))$. This scaling relation characterizes both  fat tails
and volatility clustering exhibited by financial time series. The
fat tails are described by a universal scaling function $f(z)$.
The functional form of the scaling factor $w(r_p)$, on the other
hand, contains the essential information about volatility
clustering on the time scale under consideration. The scaling
factors we obtain from the stock market data allow us to identify
regimes of high and low volatility clustering. We also present a
simple phenomenological model which captures some of the key
empirical features.

The key ``stylized facts'' about asset returns include the
following: The unconditional distribution of returns shows a
scaling form (fat tail). The distribution of returns $r$ in a
given time interval (defined as the change in the logarithm of the
price normalized by a time-averaged volatility) is found to be a
power law $P(|r|
> x) \sim x^{-\eta_r}$ with the exponent $\eta_r \sim 3$ for U.S.
stock markets\cite{gopikrishnan,gabaix}, well outside the L\'{e}vy
stable range of 0 to 2. This functional form holds for a range of
intervals from minutes to several days while for larger times  the
distribution of the returns is consistent with a slow crossover to
a normal distribution. Another key fact is the existence of
volatility clustering in financial time series that is by now well
established \cite{benoit,Fama,Engle,Bollerslev,intro4}; it can be
seen, for example, in the absolute value of the return $|r|$,
which shows positive serial correlation over long lags (the Taylor
effect \cite{Taylor}). This long memory in the autocorrelation in
absolute returns, on a time scale of more than a month, stands in
contrast to the short-time correlations of asset returns
themselves. Fat tails have been the subject of intense
investigation theoretically from Mandelbrot's  pioneering early
work\cite{benoit} using stable distributions to
 agent-based models of Bak {\em et al.}\cite{bak} and Lux\cite{lux0,lux}
 (See Ref.~\cite{LeBaron} for a survey of research on agent based models used in finance).
 The key problem is to elucidate the
nature of the underlying stochastic process that gives rise to
both volatility clustering and the power-law (fat) tails in the
distribution of asset returns.

\section{Conditional Probability and Scaling Form}
 In an effort to seek a
direct quantitative characterization of clustering  we consider
$P(r|r_p)$, the probability of the return $r$ in a time interval
of duration $T$, conditional on  the absolute value of the return
$r_p$ in the previous interval of the same duration. (We emphasize
that the probability is not conditioned on the value of the return
at an instant.) By varying $T$, we can check volatility clustering
on different time scales. There is a growing literature on
conditional measures of distribution for analyzing financial time
series (for a review, see Ref.~\cite{Malevergne} and references
therein). For example, the conditional probability of return
intervals has been used recently to study scaling and memory
effects in stock and currency data \cite{Yamasaki}.

We have analyzed both the high frequency data and daily closing
data of stock indices and individual stock prices using the
conditional probability as a probe. Here we only present results
of our analysis of high frequency data of QQQ (a stock which tracks the NASDAQ 100 index)
from 1999 to 2004 and daily closing data of the Dow Jones Industrial
Average from 1900 to 2004. We emphasize that the properties of the
financial time series we present are rather general: we have
checked that the same properties are also exhibited in other stock
indices and future data (for example, the Hang Seng Index, Russell 2000 Index, and
German government bond futures) as well as individual stocks. We have
checked, as was found in the previous studies \cite{gopikrishnan},
that the probability distribution of the returns in the time
intervals $T = 1, 2, 4, 8, 16, 32$ days for DJIA exhibits a fat
power-law tail with an exponent close to $-4$; this appears to be
true for most stock indices and individual stock data.

\begin{figure}
\includegraphics*[width=8.5cm]{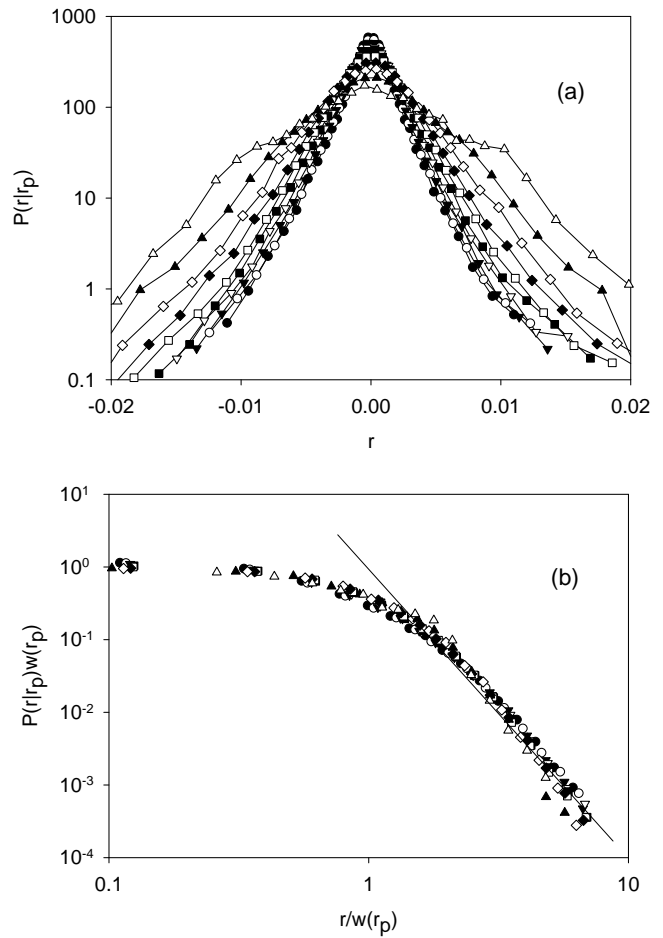}
\caption{(a) Conditional probability of return for $ T = 5$ minutes in
QQQ. Different curves correspond to 10 different absolute values of the return $r_p$
in the previous interval, which are groups of bins centered at
values ranging from  $8.4 \times 10^{-4}$ to $0.011$. The larger
the value of $r_p$, the large the width of the distribution. (b) The conditional
probability distribution of return of QQQ (shown in (a)),
when scaled by a scale factor $w(r_p)$, collapses to a universal curve. $r_p$ is
the absolute value of
the return in the previous interval. The tail of the
probability distribution can be described by a power law with the
exponent approximately equal to $-4$.}
\end{figure}

We calculate $P(r| r_p)$, by grouping the data into different bins
according to the value of $r_p$. In Figure 1(a) we display
$P(r,T|r_p, T)$ for $T=5$ minutes for different values of $r_p$. It
is clear from the figure that there is a positive correlation
between the width of $P(r| r_p)$ and $r_p$. What is more
interesting is that, when $r$ is scaled by the width of the
distribution (the standard deviation of the conditional return),
$w(r_p)$, the different curves of conditional
probability collapse to a universal curve: $P(r| r_p) =
w(r_p)^{-1} \bar{P}(r/w(r_p))$. Evidence for this is displayed in
Fig.~1(b). Note that on the time scales we have analyzed, the
probability distribution is symmetric with respect to $r$.
Consequently, in Fig.~1(b) we have only displayed the absolute value
of the return. The data collapse is good for a wide range of $T$,
and the curves display a power-law tail with  a well-defined
exponent of approximately $-4$.

We examine next the behavior of the scale factor $w(r_p)$ on
$r_p$. Fig.~2 shows a plot of the scale factor $w(r_p)$ vs. $r_p$
for different values of $T$. It can be seen from the figure that
there is a crossover value $r_c(T)$: for $r_p < r_c(T)$, $w$ is
almost constant, while for $r> r_c(T)$, $w$ increases with $r_p$.
The degree of the dependence of $w$ on $r_p$ can be taken as an
indication of strength of volatility clustering. If there is no
volatility clustering $w$ will not depend on $r_p$. Note that
there is a strong clustering at small $T$. As $T$ increases, the
strength of clustering gradually decreases, indicating a crossover
to the non-clustering regime. As $T$ increases beyond the time
scale of volatility clustering,  the clustering disappears. This
crossover can not be seen in the QQQ data as the time scales
involved are small. Our analysis of DJIA data show an indication
of such crossover at the time scale of a few months. In
 this paper, we do not separate the cases of
positive and negative returns in the previous time interval. Thus
we do not show explicitly the well-known leverage effect, first
expounded by Black \cite{Black}. We have checked that the scaling
and data collapse we obtained are equally valid when we separate
out the cases of positive and negative returns in the previous
interval. The leverage effect is reflected in the scaling factor
$w(r_p)$, which shows $w(-r_p) > w(r_p)$ for $r_p > 0$ in the real
data.

\begin{figure}
\includegraphics*[width=8.5cm]{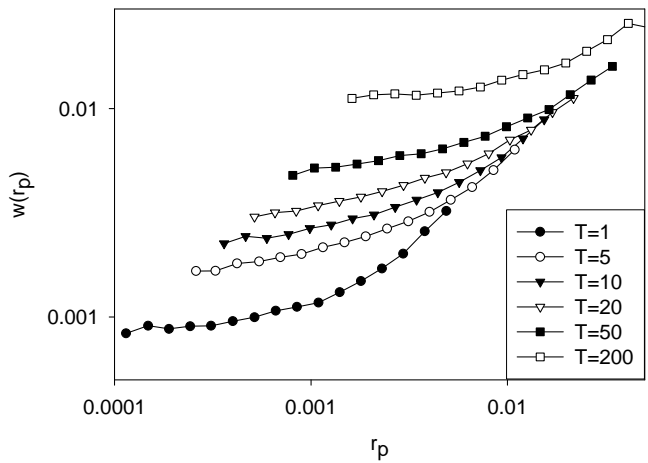}
\caption{ The scale factor $w(r_p,T)$ vs $r_p$ (the absolute value of the return
in the previous interval) for different values
of $T$, arising from analysis of QQQ data. The dependence is seen
to be almost linear for sufficiently large $r_p$}
\end{figure}

Figure~3 shows that the same scaling form is also exhibited in DJIA data. We have
checked that the data collapse extends also to
data for different values of $T$ in addition to different values
of $r_p$ displayed here.

\begin{figure}
\includegraphics*[width=8.5cm]{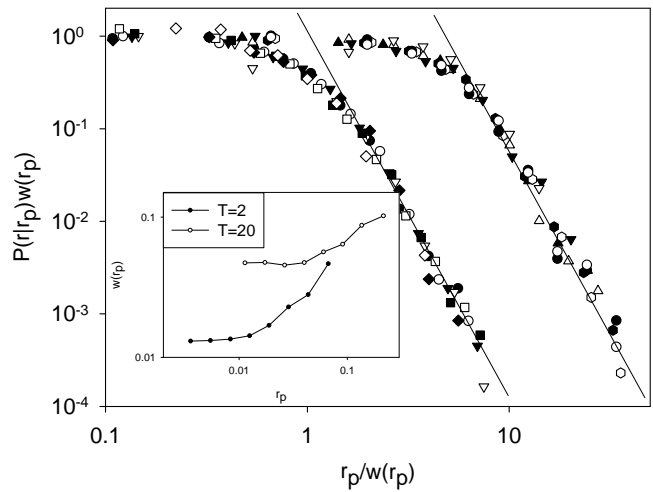}
\caption{ Conditional probability of return for $ T = 2$ and $T=20$ days in
DJIA data. The data corresponding to $T=20$ have been shifted to the right for easy viewing. Different curves correspond to 8 different absolute values of the return
$r_p$ in the previous interval. The inset shows the dependence of the width $w(r_p)$ on $r_p$.
The tail of the probability distribution can be described by a power law with the exponent
approximately equal to $-4$.}
\end{figure}

 The  data
collapse we have displayed for different $r_p$ and different $T$,
the power-law behavior including the value of the exponent, and
the behavior of the scale factor which encapsulates features of
volatility clustering are the same across data from several other
stock indices listed earlier and individual stocks. This empirical universality
can be stated as
\begin{equation}
P(r|r_p) = \frac{1}{w(r_p, T)}f(r/w(r_p, T)).
\end{equation}
Here $f(z)$ is a universal function describing the universal fat
tail in the distribution. $f(z)$ satisfies $f(z)\rightarrow$
constant as $z\rightarrow 0$, $f(z) \rightarrow 1/z^4$ as $z \gg
1$, and $\int^{\infty}_{0}f(z)dz = 1$. The dependence of $w(r_p,
T)$ on $r_p$, on the other hand, describes the volatility
clustering at the time scale $T$. If $w(r_p)$ is a constant
(independent of $r_p$), then $P(r|r_p)$ does not depend on $r_p$,
and there is no volatility correlation or clustering. The
conditional probability distribution contains information about
the conditional average of the moments $<r^q>_{r_p}$ of the
distribution as well as various volatility correlation functions
such as $<r^2r_p^2>$. Given the scaling form we can evaluate these
averages and correlation functions in terms of $w(r_p)$, which is
itself given by $w(r_p) = \sqrt{<r^2>_{r_p}}$. In particular, we
have the moments of the conditional probability distribution given
by $M_q(r_p) = <r^q>_{r_p} =C_q w^q(r_p)$ ($C_q$ is a universal
constant) and $<r_p^2r^2> = \int dr_pQ(r_p)w^2(r_p)r^2_p$, where
$Q(r)$ is the unconditional probability distribution of the
return.  We believe that this scaling form  provides a new and
rather complete measure of volatility clustering.

\section{Model and Discussion}

In the following we will provide the outline of a model that
captures the key features exhibited in the conditional probability
distribution of stock market data. In a stochastic volatility
model, the one-step asset return at time $t$ is written as
$\Delta_t = \delta_t z_t$, where $z_t$ is a Gaussian random
variable with zero mean and unit variance and $\delta_t$ is
magnitude of the price change. For the relatively short time
scales we are interested in we have set the intrinsic growth rate
to zero.  The distribution of $r$ depends on the dynamics of
$\delta_t$: Slow changes in $\delta_i$ lead to volatility
clustering.

There exist a few classes of volatility models that have been used
to describe the dynamics of $\delta_t$. These include the widely
used models based on GARCH-like processes \cite{Engle}, and more
recently, the models based on a multifractal random walk (MRW)
\cite{MRW} that will be discussed later.  In our model, the
dynamics of $\delta(t)$ is specified via  the random variable
$n(t)$, with $\delta(t) = \delta_0 \gamma^{n(t)}$. In order to
describe both the behavior of probability distributions and
temporal correlations we have devised the following model for the
evolution of $n(t)$. The time evolution of the variable $n$ is
assumed to be independent of the change in $S(t)$ and  $n(t)$
executes a random walk with reflecting boundaries: We enforce the
condition $n(t) \ge 0$; thus $\delta_0$ is the minimum value of
$\delta(t)$. An upper bound in $n$, $n_{max}$, can also be
incorporated without affecting the scaling behavior of the model.
We typically choose $\gamma^{n_{max}} \sim 30$. The change in
$n(t)$, $ \delta n(t)$ is given by
\begin{eqnarray}
\label{dn}
\delta n(t)\,&=&\,\eta_t\,+\,\alpha\{\sum_{i=1}^{N_c}
[K(i+1)-K(i)]\eta_{t-i} +K(1)\eta_t \nonumber \\
    & &  -K(N_c+1)\eta_{t-N_c} \}\,-\,\beta\overline{\eta}\,.
\end{eqnarray}
In the preceding $\{\eta_j\}$ are independent random variables
that assume the value $+1$ with probability $p$ and $-1$ with
probability $1-p$. This asymmetry builds in the tendency to
decrease the volatility. The mean value of $\eta_i$, $2p-1<0$ is
denoted by $\overline{\eta}$. We comment on the implications of
the different terms next.

We focus on the limit  $\alpha=0$ and $\beta=0$ first since it is
amenable to analytic investigation; this model is related to a
model discussed in Ref.~\cite{chen}. Note that this limit already
builds in volatility clustering  as it takes many steps to change
$n(t)$ significantly. It is easy to show that, the steady-state
probability distribution of $n$ is given by $P(n) =
(1-e^{-\lambda})e^{-\lambda n} \sim \lambda e^{-\lambda n}$, where
$\lambda = \ln((1-p)/p)$. The distribution of $\delta(t)$ is then
given by a power-law, $P(\delta) \sim \delta^{-\lambda/\ln \gamma
-1}$. This mechanism for  generating a power-law distribution was
first noted by Herbert Simon \cite{simon} in 1955.  We have
studied this limiting case of the model numerically and find that
many features of the conditional probability distribution
exhibited by the real data including the power law and  scaling
behaviors are reproduced.  We can show analytically that the
conditional probability distribution exhibits scaling collapse,
and that scale-invariant behavior with a power law tail (with the
exponent $-4$ if we choose $p=1/(1+\gamma^2)$) exists for $r
>\sigma_c$, where $\sigma_c = \sigma_0 \gamma^{(T/\delta t)
(1-2p)}$. The numerical data in fact show a somewhat larger range
of power-law behavior. The re-scaling factor required for data
collapse is simply proportional to $r_p$ from our analysis, as we
have observed from the real data and from numerical simulations of
model when $r_p$ is not too small. The simple limit captures
important features of volatility clustering reflected in
conditional probability distributions.

The second term in Eq.~(\ref{dn}) is based on the multifractal random walk model
that builds in long-time correlations via a logarithmic decay of
the log volatility correlation $\langle \log\vert r(t+\tau)\vert
\log \vert r(t)\vert\rangle$. This term allows us to reproduce the
more subtle temporal autocorrelation behavior observed in the data
and follows the implementation in Ref.~\cite{Sornette}. The
long-term memory effects are incorporated by  making the change in
$n(t)$ depend on the steps $\eta_{t-i}$ at earlier times with a
kernel given by $K(i)\,=\,1/\sqrt{i}$ (this corresponds to the MRW part of the
model given by $n(t) = \alpha\sum_{i=1}^{N_c}K(i)\eta_{t-i}$) and
allowing memory up to $N_c$ time steps, chosen to be $1000$ in our
simulations. The final term allows us to control the rate of drift
to lower values of $n$. We have simulated this model with
$\alpha_0=0.1$ ($\alpha=\alpha_0/\ln(\gamma)$)
and $\beta\approx 1.3$ for $\gamma=1.05$ and
displayed the results for $P(r\vert r_p)$ in Figure 4. The model
with the stated parameters reproduces the fat tail in the
unconditional probability distribution for $r$ observed in the
data.  The non-universal scale factor $w(r_p)$ is similar to those
found from our empirical analysis. We have also checked that this
model retains the same temporal behavior in the log-volatility
correlation exhibited by the pure MRW model. Thus the model we
have investigated is capable of reproducing both probability
distributions (conditional and unconditional) and temporal
autocorrelations. We note in passing that the model as it stands
cannot be used to study the leverage effect; however, it can be
modified to do so.

\begin{figure}
\includegraphics*[width=8.5cm]{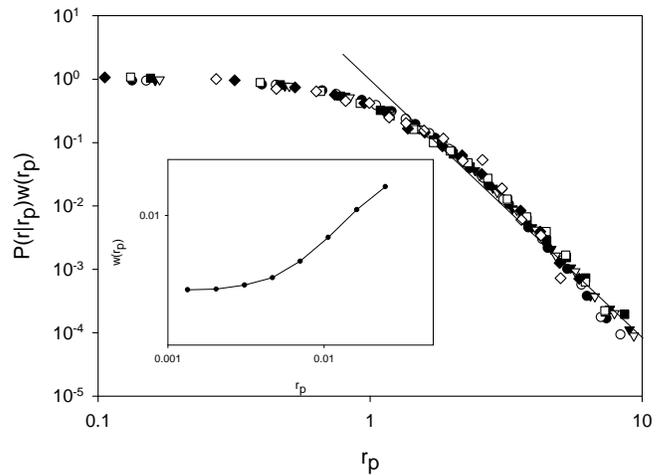}
\caption{ The scaled conditional probability distributions of
return for the mixed model given by Eq.~(2) with $\gamma=1.05$, $\alpha_0=0.1$, and $\beta=1.3$.
The time lag is $T=10$.  The curves,
corresponding to different absolute values of the return $r_p$ in the previous interval
collapse on to a universal curve when scaled by a scale
factor $w(r_p)$.  The tail of the probability distribution is again described by a power law
with the exponent equal to $-4$. The inset shows the dependence of $w(r_p)$ on $r_p$. }
\end{figure}

In summary, we have proposed a direct measure of volatility
clustering in financial time series based on the {\em conditional}
probability distribution of asset returns over a time period given
the return over the previous time period.  We discovered that the
conditional probability of stock market data can be well described
by a scaling relation, which reflects both fat tails and
volatility clustering of the financial time series. 
In particular,
the strength of volatility clustering is reflected in the
functional form of the scaling factor $w(r_p)$. By extracting
$w(r_p)$ from market data, we are able to estimate the future
volatility over a time period, given the return in the previous
period. This may be useful in modelling financial transactions
including option pricing, portfolio and risk management; all these
depend crucially on volatility estimation. The clustering of
activities and fat tails in the associated distribution are very
common in the dynamics of many social \cite{Barabasi} and natural
phenomena (e.g. earthquake clustering
\cite{Kagan}). The conditional
probability measure we have presented in this paper may serve as a
useful tool for characterizing other clustering phenomena.

This work was supported by the National University of Singapore research
grant R-151-000-032-112.


\begin{thebibliography}{0}

\bibitem{Engle} R. F. Engle, Econometrica {\bf 50}, 987 (1982).
\bibitem{benoit} B. Mandelbrot, J. Business {\bf 36}, 394 (1963).
\bibitem{Fama} E. F. Fama, J. Business {\bf 38}, 34 (1965).
\bibitem{intro1} R. Mantegna and H. E. Stanley {\it An introduction to econophysics} (Cambridge University Press, Cambridge, 1999).
\bibitem{intro2} J.-P. Bouchaud and M. Potters {\it Theory of financial risks: from statistical physics to risk management} (Cambridge University Press, Cambridge, 2000).
\bibitem{intro3} R. Cont {\it Quantitative Finance} {\bf 1}, 223 (2001).
\bibitem{intro4} R. F. Engle and A. J. Patton Quantitative Finance {\bf 1} 237 (2001).
\bibitem{Plerou} V. Plerou, P. Gopikrishnan, L.A.N. Amaral, X. Gabaix, and H.E. Stanley, Phys. Rev. E {\bf 62}, R3023 (2000).
\bibitem{gopikrishnan} P. Gopikrishnan, M. Meyer,  L. A. N. Amaral, and  H. E. Stanley, Euro. Phys. J. B {\bf 3}, 139 (1998).
\bibitem{gabaix}  X. Gabaix,  P. Gopikrishnan, V. Plerou, and  H. E. Stanley, Nature {\bf 423}, 267 (2003).
\bibitem{Bollerslev}  T. Bollerslev, Journal of Econometrics {\bf 31}, 307 (1986).
\bibitem{Taylor} S. Taylor, {\it Modelling the financial time series}
    (John Wiley, New York 1986).
\bibitem{bak}  P. Bak,  M. Paczuski, and  M. Shubik, Physica A {\bf 246}, 430 (1997).
\bibitem{lux0} T. Lux  and M. Marchesi, Nature {\bf 397}, 498 (1999).
\bibitem{lux} T. Lux, Journal of  Economic Behavior and Organization {\bf 33}, 143 (1998).
\bibitem{simon}  H. Simon, Biometrika {\bf 42}, 425 (1955).
\bibitem{LeBaron} B. LeBaron, in {\it Handbook of Computational Economics, Vol. 2: Agent-Based Computational Economics}, eds. L. Tesfatsion  \& K.L. Judd  (North-Holland, Amsterdam), Chapter 9 (2006).
\bibitem{Malevergne}  Y. Malevergne and  D. Sornette {\it Extreme Financial Risks (from dependence to risk management)} (Springer, Heidelberg 2005).
\bibitem{Black} F. Black, Proc. Bus. Econ. Statist. Sect. Am. Statist. Assoc., 177 (1976).
\bibitem{MRW}  E. Bacry,  J. Delour, and  J. F. Muzy, Phys. Rev. E {\bf 64}, 026103 (2001).
\bibitem{Sornette}  D. Sornette,  Y. Malevergne,  J. F. Muzy, Risk Magazine {\bf 16}(2), 67 (2003).
\bibitem{Yamasaki}  K. Yamasaki,  L. Muchnik,  S. Havlin,  A. Bunde, and  H. E. Stanley, Proc. Natl. Acad. Sci. USA {\bf 102}, 9424 (2005).
\bibitem{chen} K. Chen and C. Jayaprakash, Physica A {\bf 324}, 258 (2003).
\bibitem{Barabasi}  A.-L. Barabasi,  Nature {\bf 435}, 207 (2005).
\bibitem{Kagan} Y. Y. Kagan and  D. D. Jackson, Geophys. J. Int. {\bf 104}, 117 (1991).
\end{thebibliography}
\end{document}